# Optical waveguide hosting multiple exceptional points: Toward selective mode conversion

Arnab Laha,[1] Harsh K. Gandhi,[1] Sibnath Dey,[1] Sayan Bhattacherjee,[1] Abhijit Biswas,[2] and Somnath Ghosh [1, *]

[1] Department of Physics, Indian Institute of Technology Jodhpur, Rajasthan-342011, India
[2] Institute of Radiophysics and Electronics, University of Calcutta, Kolkata-700009, India
* *somiit@rediffmail.com*

**We investigate the astonishing physical aspects of Exceptional Points (EPs) in a 1D planar few-mode optical waveguide. The waveguide hosts four quasi-guided modes. Here interactions between the selected pair of modes are modulated by a spatial distribution of inhomogeneous gain-loss profile. Both the coupled pairs approach two different second order EPs in parameter plane. Considering a proper parametric loop to encircle the identified EPs simultaneously, we establish a specific topological feature where one round encirclement in parameter space yields the switching between the propagation constants ($\beta$) of the corresponding pairs of couple modes in complex $\beta$-plane. Choosing two different patterns of the parametric loop, we establish the immutable topology in $\beta$-switching phenomena. This robust mode conversion scheme shall provide a platform to realize selective mode switching devices or optical-mode converters.**

1. INTRODUCTION:

Non-Hermitian formalism in quantum mechanics provides some revolutionized tools to understand various open quantum-inspired or wave-based photonic structures. Due to formal equivalence of the Schrödinger equation with the Helmholtz equation under paraxial approximation, the potential function $V(x)$ in such open photonic systems can easily be realized with the refractive index profile $n(x)$; where the optical gain-loss can be integrated to attain system non-Hermiticity. One of the distinct non-Hermitian features is phenomena of the presence of exceptional points (EPs) which has a growing interest for research in topological photonics recently. An EP can be considered as a topological branch point singularity in at least 2D parameter space where the coupled eigenvalues and the corresponding eigenvectors of the underlying Hamiltonian simultaneously coalesce [1,2]. An EP leads to significant modifications in behaviour of corresponding coupled eigenvalues due to perturbation parameters. In parameter space with moderate encirclement around a second order EP via adiabatic variation of chosen control parameters, one of the associated eigenvalues can flip with its coupled counterpart (i.e. exchanging their positions) in eigenvalue plane [3-6] and during permutation one of the corresponding eigenstates acquires the geometric phase and interchanges with its coupled counterpart [6]. Here, an EP exhibit as second-order branch point for eigenvalues and forth-order branch point for eigenvectors [4-8]. This state flipping mechanism is called flip-of-states which leads a key role in the context of optical mode conversion.

Exotic behaviors of EPs have been extensively studied in various photonic systems like optical microcavity [3-6], waveguide [7-10], laser [11], photonic crystal [12] and also in several non-optical systems like atomic [13] and molecular spectra [14], microwave cavity [2], etc. With precise control over fabrication, EP in photonic system can offer itself as a strong competitor to meet the present-day challenges like mode conversion [4-10], unidirectional light propagation [4-6], ultrasensitive EP aided sensing [3], to name a few.

In this paper, we report a symmetric step-index planar optical waveguide that supports four quasi-guided modes. Here we introduce non-Hermiticity with a spatial distribution of transverse inhomogeneous gain-loss profile. The interaction between the supported modes are controlled by proper tuning of two independent parameters viz. gain-coefficient ($\gamma$) and loss-to-gain ratio ($\tau$). The interacting fundamental mode ($\psi_0$) and first higher order mode ($\psi_1$) approach a second order EP (say, $EP^{(1)}$) in ($\gamma, \tau$)-plane; whereas next two higher order modes i.e. $\psi_2$ and $\psi_3$ approach a different second order EP (say, $EP^{(2)}$) in the same parameter plane. Interestingly, we observe that the interaction between the pair $\psi_0$ and $\psi_1$ are not affected by $\psi_2$ and $\psi_3$ and also vice-versa which essentially reflects the fact that two identified EPs i.e. $EP^{(1)}$ and $EP^{(2)}$ are independent and noninteracting. Now, we consider two different parameter spaces individually in ($\gamma, \tau$)-plane to encircle both the EPs simultaneously and establish mode-switching between the pairs ($\psi_0$, $\psi_1$) and ($\psi_2, \psi_3$). Our proposed waveguide with rich physical aspects of EPs is efficiently able to realize selective mode switching and may provide a platform to develop optical mode converters, optical switches, etc.

## 2. DESIGN OF THE OPTICAL WAVEGUIDE:

We design a step-index planar optical waveguide, schematically shown in Fig. 1. Normalizing the operating frequency $\omega = 1$ i.e. wavelength $\lambda = 2\pi$, we set the total width of the waveguide $W = 50\lambda/\pi = 100$ (in a dimension-less unit) The waveguide occupies the region $-W/2 \leq x \leq W/2$ with a core (with in $-W/6 \leq x \leq W/6$) and a cladding (with in $W/6 \leq |x| \leq W/2$) having passive refractive indices $n_h$ and $n_l$, respectively. Here we choose $n_h = 1.5$ and $n_l = 1.46$. The parameters are chosen in such a way that the waveguide supports only four quasi-guided modes $\psi_j$ ($j = 0,1,2,3$).

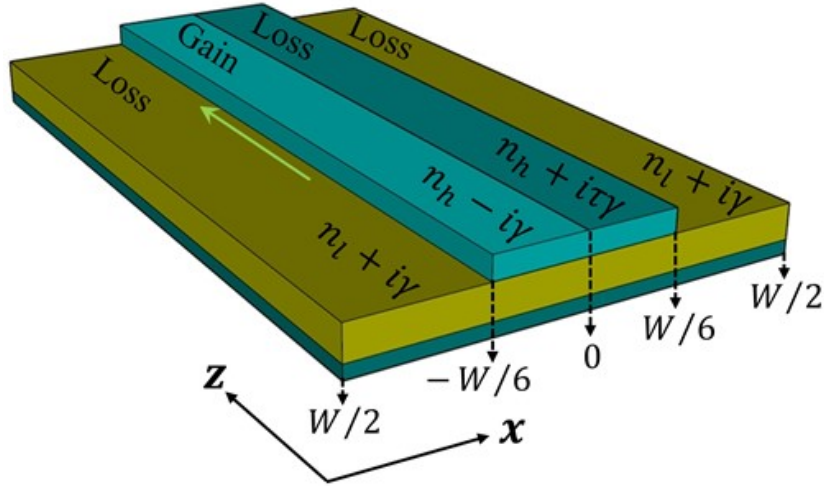

Fig. 1. Schematic diagram of the proposed waveguide with transverse $x$-direction. Here propagation is considered along the $z$-direction.

The interactions between the consecutive modes with one-to-one coupling restriction is controlled by the system non-Hermiticity which is achieved by a spatial distribution of transverse inhomogeneous gain-loss profile. Here the distribution of gain-loss profile is characterized by two independent parameters viz. gain-coefficient ($\gamma$) and loss-to-gain ratio ($\tau$). Here, $\tau$ essentially represent the inhomogeneity in gain-loss profile in the waveguide core. Thus, the overall index profile can be written as

$$n(x) = \begin{cases} n_h - i\gamma & -W/6 \leq x \leq 0 \\ n_h + i\tau\gamma & 0 \leq x \leq W/6 \\ n_l + i\gamma & W/6 \leq |x| \leq W/2 \end{cases} \quad (1)$$

The propagation constants ($\beta_j; j = 0,1,2,3$) of the supported modes ($\psi_j; j = 0,1,2,3$) have been calculated from the equivalent form of the Maxwell's equation

$$[\partial_x^2 + n^2(x)\omega^2 - \beta^2]\psi(x) = 0 \quad (2)$$

using $n(x)$ as given in Eq. 1. Due to small index difference between core and cladding the scalar modal equation given by Eq. 2 is valid as long as the supported modes are guided in the core region.

## 3. NUMERICAL RESULTS:

### A. Encounter of Exceptional Points (EPs)-

Considering the optimized operating parameters, in this section we investigate the interactions between the supported modes with introduction of unbalanced gain-loss profile. Using the concept of avoided resonance crossing (ARC) [1-2, 4-8] we identify two second order EPs in ($\gamma, \tau$)-plane when the complex $\beta$-valus associated with coupled eigenstates coalesce in complex $\beta$-plane.

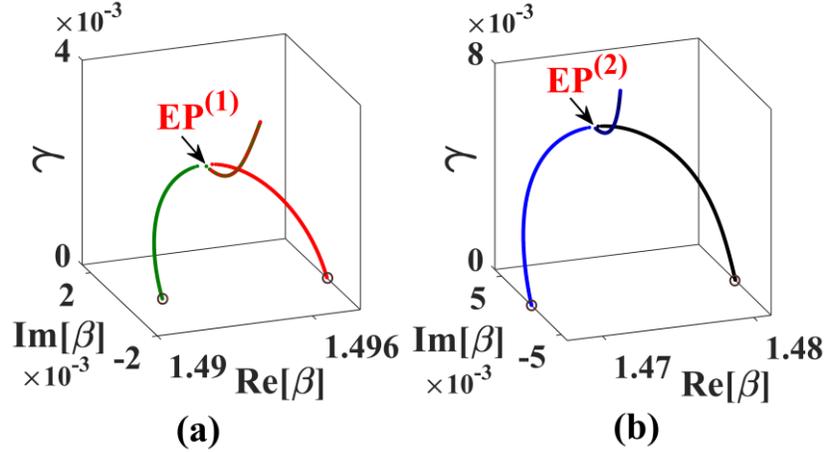

Fig. 2. Encounter of EPs: (a) Dynamics of the complex $\beta_0$ (dotted red curve) and $\beta_1$ (dotted green curve) with respect to $\gamma$ for a chosen $\tau = 3.365$ showing coalescence at EP$^{(1)}$ near $\gamma = 0.0017$. (b) Dynamics of the complex $\beta_2$ (dotted black curve) and $\beta_3$ (dotted blue curve) with respect to $\gamma$ for a chosen $\tau = 3.452$ showing coalescence at EP$^{(2)}$ near $\gamma = 0.0051$. The brown circular markers in both (a) and (b) represent the initial locations of each $\beta$-values.

At first choosing $\psi_0$ and $\psi_1$, we investigate the ARC phenomena between $\beta_0$ and $\beta_1$ with crossing/anticrossing of their real and imaginary parts in a specified $\gamma$-range from 0 to 0.01, however, for different $\tau$-values. Investigating several cases, we observe that for $\tau = 3.365$, $\beta_0$ and $\beta_1$ coalesce near $\gamma = 0.0017$ as can be seen in Fig. 2(a). Thus, we numerically detect the position of an EP at ($\gamma_{EP} = 0.0017$; $\tau = 3.365$) which has been denoted as EP$^{(1)}$. Similarly, we study the interaction between $\psi_2$ and $\psi_3$ and identify a different EP at ($\gamma_{EP} = 0.0051; \tau = 3.452$) that has been denoted as EP$^{(2)}$. As can be seen in Fig. 2(b), $\beta_2$ and $\beta_3$ coalesce at EP$^{(2)}$.

## B. Effect of parametric encirclements around EPs-

To investigate the EP-encirclement scheme, we consider a closed circular parametric loop in ($\gamma, \tau$)-plane following the coupled equation

$$\gamma(\phi) = \gamma_0(1 + a \cos \phi) \text{ and } \tau(\phi) = \tau_0(1 + a \sin \phi) \tag{3}$$

to enclose both the identified EPs. Here ($\gamma_0, \tau_0$) represent the center of the circle. $a$ ($\in (0,1]$) is a characteristics parameter and $\phi$ ($\in [0,2\pi]$) is tunable angle.

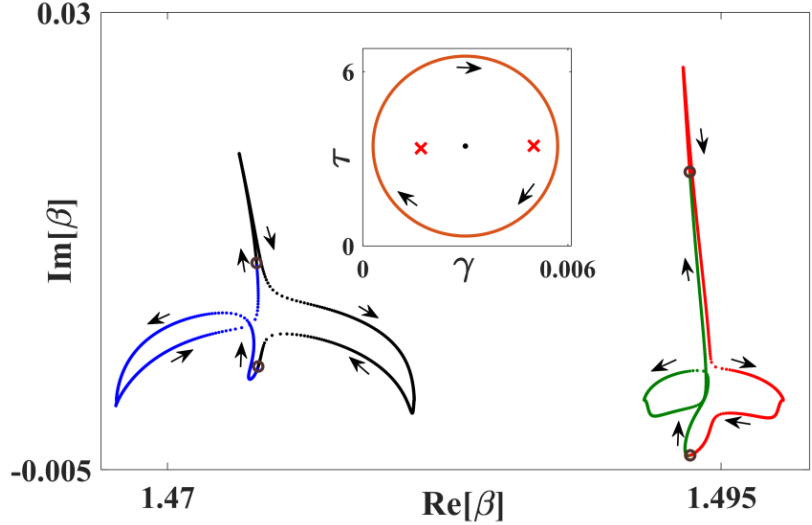

Fig. 3. Trajectories of the propagation constants of the respective coupled modes in complex $\beta$-plane corresponding to both the EPs (marked by red crosses in the inset) for a common encircling process following Eq. 3 as shown in inset. Each point on the four $\beta$-trajectories in the complex $\beta$-plane is generated by each point evolution in the parametric loop in ($\gamma, \tau$)-plane. Here arrows indicate direction of progressions. The brown circular markers represent the initial locations of each $\beta$-values.

Now we judiciously set the center at ($\gamma_0 = 0.003; \tau_0 = 3.44$) (indicated by black dot in the inset of Fig. 3) and $a = 0.9$ to enclose both EP$^{(1)}$ and EP$^{(2)}$. The parameter space is shown in the inset of Fig. 3. Now, following a quasi-static parametric variation along the closed circular loop in ($\gamma, \tau$)-plane we observe the permutation between $\beta_0$ and $\beta_1$, and also between $\beta_2$ and $\beta_3$ in complex $\beta$-plane. Here $\beta_0$ is going to the initial position of $\beta_1$ as shown by dotted red trajectory in Fig. 3 and simultaneously $\beta_1$ is going to initial position of $\beta_0$ as shown by dotted green trajectory in Fig. 3. Similarly, following the same encirclement process $\beta_2$ and $\beta_3$ exchange their initial positions as shown by dotted black and blue trajectories respectively, and form a closed loop in complex $\beta$-plane. Here brown circular markers indicate the initial positions of $\beta$-values. In this context, if we choose such an $a$-value for which one the EPs goes outside the parametric loop then $\beta$-values of the corresponding eigenstates ignore permutation and make individual loops in complex $\beta$-plane.

Now, we consider a different parametric contour in $(\gamma, \tau)$-plane having a different shape which is given by the coupled equations

$$\gamma(\phi) = \gamma_0 \sin\left(\frac{\phi}{2}\right); \; \gamma_0 > \gamma_{EP} \text{ and } \tau(\phi) = \tau_{EP} + a \sin \phi \tag{4}$$

For $\gamma_0 < \gamma_{EP}$, the parameter space does not enclose the EP. Eq. 4 is necessary for device application [7-8] beyond the chosen parameter space given by Eq. 3 because Eq. 4 considers the situation $\gamma = 0$ for both $\phi = 0$ and $\phi = 2\pi$. Thus, one should find clean passive modes at the end of the encirclement.

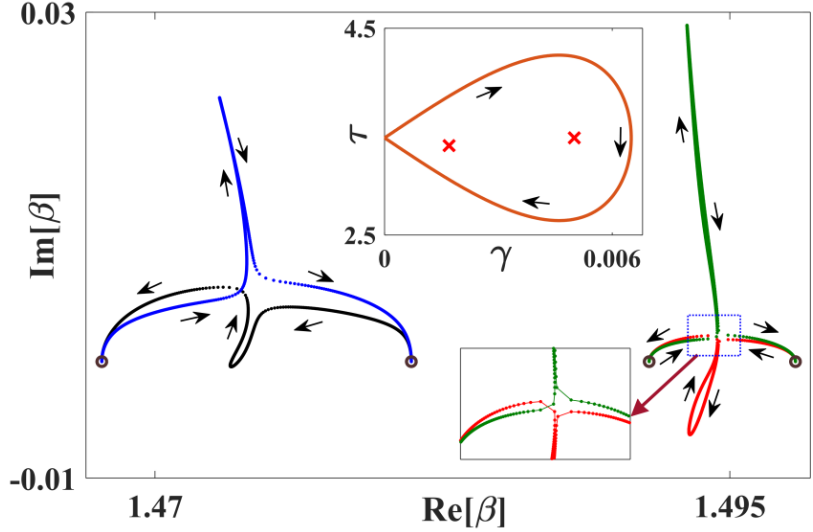

Fig. 4. Trajectories of the propagation constants of the respective coupled modes in complex $\beta$-plane corresponding to both the EPs (marked by red crosses in the inset) for a common encircling process following Eq. 4 as shown in inset. Here arrows indicate direction of progressions. The brown circular markers represent the initial locations of each $\beta$-values. In the trajectories of $\beta_0$ and $\beta_1$, the dotted rectangular portion is zoomed in the inset for clear view.

Now, choosing $\gamma_0 = 0.008$ and $a = 0.9$, we set a parameter space essentially to enclose both the EPs which is shown in the inset of Fig. 4. The corresponding trajectories of the propagation constants are shown in Fig. 4. Here also we observe that for one round parametric encirclement the coupled pairs $(\beta_0, \beta_1)$ and $(\beta_2, \beta_3)$ mutually exchange their initial positions in a generic fashion as we have already seen in Fig. 3. In this context, the modal propagations and corresponding asymmetric mode conversion [7-9] will be further reported.

The results shown in Figs. 3 and 4 reflect the fact that the $\beta$-swtching phenomena is topological because it is independent of the shape of the parameter space. It is evident that the trajectory of $\beta_0$ and $\beta_1$ is unaffected from the influence two nearby modes $\beta_2$ and $\beta_3$, and also vice-versa. Here two identified EPs are noninteracting. Thus, we can selectively convert one mode to another only by encircling the respective EP ignoring the annoying influences of nearby modes.

## 4. CONCLUSION:

In summary, we report a four-mode optical waveguide with transverse distribution of inhomogeneous gain-loss profile. The waveguide hosts two non-interacting second order EPs. The consecutive modes are coupled with one-to-one coupling restriction and are analytically connected by two identified EPs. The waveguide supports parameter spaces to encircle both the EPs. As long as the EPs are encircled in the parameter space the $\beta$-switching phenomena is omnipresent and also independent of the shape of the parametric loop. These phenomena are also unaffected by any nearby states. The proposed waveguide with our exclusive scheme should be useful for implementation of selective mode switching in photonic devices.

ACKNOWLEDGMENTS:

AL and SG acknowledge the financial support from the Science and Engineering research Board (SERB) under Early Career Research Grant [ECR/2017/000491], and the Department of Science and Technology (DST) under INSPIRE Faculty Fellow Grant [IFA-12, PH-23], Ministry of Science and Technology, India.